\documentstyle[aps,multicol,amssymb,epsfig]{revtex}
\def\one{{\mathchoice {\rm 1\mskip-4mu l} {\rm 1\mskip-4mu l} {\rm
1\mskip-4.5mu l} {\rm 1\mskip-5mu l}}}
\newcommand{\slb}[2]{{#1}^{#2}}
\begin{document}
\draft
\title{The quantum dynamics of two coupled qubits}
\author{G.~J.~Milburn$^1$, R.~Laflamme$^2$,
	B.~C.~Sanders$^3$,  E.~Knill$^2$}
\address{
$^1$~Centre for Quantum Computer Technology, The University of
Queensland, Queensland 4072 Australia.\\
$^2$~Los Alamos National Laboratory, Los Alamos, New Mexico, USA
\\
$^3$~Department of Physics, Macquarie University,
	Sydney, New South Wales 2109, Australia
 }
\date{\today}
\maketitle
\begin{abstract}
We investigate the difference between classical and quantum dynamics
of coupled magnetic dipoles. We prove that in general the dynamics of
the classical interaction Hamiltonian differs from the corresponding
quantum model, regardless of the initial state. The difference appears
as non positive--definite diffusion terms in the quantum evolution
equation of an appropriate positive phase--space probability density. Thus,
it is not possible to express the dynamics in terms of a convolution
of a positive transition probability function and the initial
condition as can be done in the classical case. We conclude that the
dynamics is a quantum element of NMR quantum information processing.
There are two limits where our quantum evolution coincide with the
classical one: the short time limit before spin--spin interaction sets
in and the long time limit when phase diffusion is incorporated.
\end{abstract}

\pacs{03.67.Hk}
\begin{multicols}{2}
\section{Introduction}

Recent work in quantum information theory has suggested that quantum
computers are more powerful than their classical
counterparts\cite{shor:qc1994a,grover:qc1995a,feynman:qc1984a,deutsch:qc1985a,lloyd:qc1996a,zalka:qc1996b}.
In quantum communication, there already exist algorithms which have
been proved to out-perform their classical counterparts
\cite{buhrman:qc1998a,raz:qc1999a}. The situation in computation is not
as clear: we know of problems which have quantum algorithms that are
exponentially faster but only when compared to the {\it known}
classical
ones\cite{shor:qc1994a,feynman:qc1984a,lloyd:qc1996a,zalka:qc1996b,wiesner:qc1996a}
(not the optimum ones). The special power of quantum computers is only
a conjecture as we have no proofs that we cannot simulate efficiently
quantum systems using classical computers.  Although most physicists
would believe this efficient simulation to be impossible, this is at
the foundation of the distinction between classical and quantum
computation.

If quantum computers are indeed more powerful than classical ones,
could we pinpoint the origin of this power to one or a few elements in
the quantum mechanical theory?  In the `folklore', the power of
quantum computation has been attributed to entanglement.  In quantum
computers where the initial states are pure it has been
claimed\cite{braunstein:qc1999a,ekert:qc1998b} that the presence of
entanglement distinguishes quantum and classical algorithms.  Indeed
the evolution of a quantum system starting in a pure state and
evolving unitarily without entanglement (which could occur if there 
were no interactions between the components) can be efficiently simulated
by a classical system; on the other hand, a classical simulation of generic quantum evolution
for a pure state has no known efficient algorithm.
At the basis of this argument is the
ability to efficiently simulate a system by a classical computer.

The argument employed in Ref.~\cite{ekert:qc1998b}, which usues the existence of 
an efficient classical simulation in the absence of entanglement, does not
carry through when the initial state is mixed, (that is not a pure state).
That is, for some
highly mixed state\cite{mixed-state} such as the state of nuclear spins
present in liquid--state NMR, we do not know how to efficiently simulate
the evolution of the system on a classical computer.  For these
states, the density matrix can be represented as a sum of separable
states with positive coefficient (to be interpreted classically as
probabilities to be in the respective states).  However, under generic
unitary evolutions, the choice of separable states must change.  No efficient algorithms exist to relate
the initial separable states to the final ones for an increasing number of spins.

We do not yet have a generic quantitative measure for entanglement,
although we do have a measure for absence of entanglement.  Pure states
are defined as being separable, or non-entangled, if they can be
expressed as products of subsystem (such as qubit) wave functions.  For
mixed states, this notion is generalized to the existence of at least
one expansion of the state in terms of separable pure states with
positive coefficients.  Thus an equal mixture of the maximally
entangled state of two spins (Bell states) does not contain any
entanglement because this density matrix can be re-expressed in terms
of separable states (for example the computational states with equal
probability).  The separable states of spin--half systems could be
described at a given time as a probability distribution of a set of
classical tops.

The notion of entanglement for mixed states has been developed in the
context of quantum communication.  One definition corresponds to the
number of maximally entangled states that can be extracted 
ensemble of these states\cite{Bennett96}.  But a computation is inherently a dynamical process,
and we do not know in general how to describe the evolution of one
mixed separable state to another using an efficient classical
description, in contradistinction to the pure state case.  The quantum
device can thus provide some information more efficiently that a
classical device could.

If mixed states are used as initial states of a quantum computer,
entanglement does not seem to play the essential role in distinguishing
quantum and classical algorithms as it might if we used pure states.  A
particular example of a quantum algorithm without a known efficient
classical counterpart is the one given in Ref.~\cite{knill:qc1998c}.
The algorithm gives the distribution of eigenvalues of a quantum
Hamiltonian.  It uses as an input state an extremely mixed state, one
with a single qubit in a pure state (a pseudo-pure state
\cite{cory:qc1996a} would also do), and all other qubits are maximally mixed.  This
algorithm uses extremely mixed states but can still outperform known
classical algorithms.  For  a small number of qubits there will
definitely be no entanglement if a pseudo-pure state of the first bit
is used; as we increase the number of bits there may or may not be
entanglement. But even if it happens at the $n^{\rm th}$ qubit, the
algorithm will not go through any phase transition; thus, it would be
meaningless to refer to the algorithm as classical before the presence
of entanglement and as quantum afterwards.  What distinguishes this
algorithm from the classical analog is that the rules for transforming
the density matrix are the quantum rules, and we do not
know how to efficiently
simulate them by the classical rules.

The algorithm in Ref.~\cite{knill:qc1998c} is especially relevant in
the context of recent discussions of experiments in quantum information
processing using liquid--state NMR technology
\cite{braunstein:qc1999a,schack:qc1999a}.  The algorithm in
Ref.~\cite{knill:qc1998c} could be implemented in liquid--state NMR.
The authors of \cite{braunstein:qc1999a} commented: ``The results in
this Letter suggest that current NMR experiments are not true quantum
computations, since no entanglement appears in the physical states at
any stage.'' This statement makes the assumption that entanglement is
the necessary element of quantum computation following the suggestion
in \cite{ekert:qc1998b}. In the same paper\cite{braunstein:qc1999a} 
however it is recognised that 
it may not be so easy to separate quantum dynamics and entanglement when 
trying to pinpoint the power of quantum coputation; 
"The results in this Letter suggest that current NMR experiments are not
true quantum computations, since no entanglement appears in the physical
states at any stage.  We stress, however, that we have not proved this
suggestion, since we would need to analyze the power of general unitary
operations in their action on separable states.  To reach a firm
conclusion, much more needs to be understood about what it means for a
computation to be a ``quantum'' computation."  However
 the claim that the evolution of unentangled pure states can be
efficiently simulated by a classical computer \cite{ekert:qc1998b} does
not carry through to mixed states.  The power of quantum computation can
come from properties of the dynamics, not the state. This was also recognised by 
Schack and Caves\cite{schack:qc1999a}. Indeed, the real
origin of the criteria in \cite{ekert:qc1998b} is the dynamical
evolution of the system not the state itself.  If we apply the type of
unitary transformation used in \cite{ekert:qc1998b} to a highly mixed
state (so that entanglement might not appear), it is as hard to
simulate on a classical computer as when the state is initially pure.
This point was not considered in
Ref.~\cite{braunstein:qc1999a}.

Recognizing this fact, Schack and Caves attempted to explain some
liquid--state NMR experiments using classical dynamics, without success
\cite{fitzgerald:qc2000}. They did not derive an equation of motion for
the behavior of the spins but rather provided a model which described
the effect of ``gates'' on the states. Their model did predict an
exponential decay of the signal as a function of the number of gates
going as $(1+2^{2n-1})^{-g}$ for $n$ the number of qubits in the
experiment and $g$ the number of gates.  Even in the three qubit
experiments they commented upon, the gates defined by Schack and Caves
are not the physical gates implemented in the experiment as this would
have ruled out their model.  In the seven--qubit experiment of
Ref.~\cite{knill:qc1999a}, the model predicted a
decrease in the signal--to--noise ratio of $\sim 10^{40}$ smaller than
observed: a number which unquestionably rules out the model.
Discrepancies of their theory with older NMR experiments, although not called quantum
computing at the time, are plentiful in the NMR literature, and the
readers can review experiments performed twenty years
ago\cite{warren:qc1979a,weitekamp:qc1982a}.

What does make the quantum dynamics so hard to simulate? Could there
be other classical models which explain NMR experiments?  Can we
understand the origin of the discrepancy between classical and quantum
evolution?  In this paper we compare the evolution of two coupled spin--half
particles under quantum and classical evolution.  The work of the last
fifty years in NMR shows the consistency of experimental results with
quantum mechanics and the failure to find a classical description:
``The dynamics of isolated spins can be understood in terms of the
motion of classical magnetization vectors. To describe coupled spins,
however, it is necessary to have recourse to a quantum mechanical
formalism where the state of the system is expressed by a state
function or, more generally, by a density
operator.''\cite{ernst:qc1994a} Here, we will give the explicit origin
for this difference for the simplest choice of a classical model: 
the one with the same Hamiltonian as the quantum model. In the next section we derive the
evolution equation for classical and quantum interacting spin--half particles,
explicitly demonstrating that the classical theory (with the same Hamiltonian)
 cannot reproduce the quantum equations.  We then discuss implications of these equations
and draw conclusions.

\section{Classical and quantum dynamics of coupled spin--half particles}

We investigate the equations of motion of classical and quantum
spin--half particle and show that the quantum behavior is fundamentally
different from the classical one.  We show that even in the cases where
the density is highly mixed (non-entangled) the evolution leads to
different observables quantities.  There is some ambiguity in exactly
what is meant by the classical dynamics of such a system. We must agree
on some ground rules to make a meaningful comparison.  Semiclassical
dynamics has a long history and some rules have been
established\cite{Gutzwiller}.
We will assign a classical analogue for a quantum problem by demanding
that the same functional form of the Hamiltonian be used but with the
corresponding classical phase space variables substituted for the
quantum canonical operators.

In the case of a spin--half system this would appear to present some
problems, but the situation is clearer if we always work in the
irreducible representations of the total angular momentum of the
system. There are two subspaces corresponding to total angular momentum
quantum number $s=1$ and $s=0$. The dynamics of a single spin--half
system will conserve angular momentum.  However, for two spin--half
systems, with arbitrary one-- and two-- qubit interactions, angular
momentum need not be conserved and these subspaces will become
coherently mixed.  It will suffice however to consider one particular
two qubit gate that does conserve angular momentum, which is an
important two qubit gate for quantum computing and for which the
classical and quantum dynamics are completely different except on a
short time scale. Furthermore this is precisely the two qubit gate
accessible in NMR quantum computing.

In order to produce entangled states of single spin--half systems,
a variety of possible interactions could be used; however, in NMR the natural
interaction is of the form
\begin{equation}
H_2=\frac{J}{4}\hat{\sigma}_z^{(1)}\hat{\sigma}_z^{(2)}
\label{two_ham}
\end{equation}
with~$\hbar=1$ and the subscript~$2$ indicating that the interaction is between two qubits.
The dynamics of two spin states that follows from this Hamiltonian is given
by a unitary operator $U_2(t)$ where
\begin{equation}
U_2(t)=\exp\left[-i\frac{J}{4}t\hat{\sigma}_z^{(1)}\hat{\sigma}_z^{(2)}\right]~.
\end{equation}
In addition to this two--spin unitary operator or `gate',  generic quantum computation needs
single spin dynamics. This is easily generated
by the scalar coupling of a spin--half magnetic dipole with an applied
magnetic field. The Hamiltonian for these single spin rotations for the
spin labelled $i$ is
\begin{equation}
H_1=  {\vec B}(t)\cdot{\vec \sigma}^{(i)}~,
\label{single_ham}
\end{equation}
with the subscript~$1$ indicated that the Hamiltonian applies to a single qubit, and the corresponding unitary operator is
transformation
\begin{equation}
U_1(t)=\exp\left[-i{\vec B}\cdot{\vec \sigma}^{(i)}\right]~.
\label{single_U}
\end{equation}
We can now use $U_1$ and $U_2$ to generate an entangled state. For example,
the Bell state
\begin{equation}
|\phi^+\rangle={|\downarrow\rangle_1\otimes|\downarrow\rangle_2+
|\uparrow\rangle_  1\otimes|\uparrow\rangle_2 \over\sqrt{2} }
\end{equation}
is generated from the product state
$|\downarrow\rangle_1\otimes|\downarrow\rangle_2$ by
\begin{eqnarray}
|\phi^+ \rangle
	&=& e^{-i\pi/4} e^{i(\pi/4) \slb{\sigma_x}{(2)}}
	e^{ i(\pi/4)\slb{\sigma_z}{(1)}}
	e^{-i(\pi/4)\slb{\sigma_y}{(2)}}
		\nonumber	\\	&&\times
	e^{-i(\pi/4)\slb{\sigma_z}{(1)}
	\slb{\sigma_z}{(2)}}	
	e^{ i(\pi/4)\slb{\sigma_y}{(2)}}
		\nonumber	\\	&&\times
	e^{-i(\pi/4)\slb{\sigma_y}{(1)}}
	\vert\downarrow\rangle_1\otimes|\downarrow\rangle_2
\end{eqnarray}

In liquid--state NMR we do not begin with initial pure states, but we
begin with a mixed state density operator
\begin{equation}
\rho_\epsilon={\cal Z}^{-1}e^{-\beta H}
\end{equation}
with~$H$ the system (individual nuclear spins on a single molecule)
Hamiltonian, $\beta=(k_B T^{-1}$ and
${\cal Z}=\mbox{tr}\rho_\epsilon$
\cite{ernst:qc1994a}.
At high temperature, as is the case for present--day liquid--state NMR quantum
computation
\cite{cory:qc1996a,chuang:qc1997a},
the state is very close to the identity so
\begin{equation}
\rho_\epsilon \sim\frac{(1-\epsilon)}{2^N}\one+\epsilon\rho_1
\label{thermal}
\end{equation}
where $d=2^N$ is the dimension of the Hilbert space for $N$ qubits, $\one$ is
the identity
operator in this tensor product space and $\rho_1$ is an arbitrary density
operator.
For example, in the case of a molecule with two spins and scalar coupling
we have,
\begin{equation}
H	= \hbar\omega_1\hat{\sigma}_z^{(1)}+\hbar\omega_2\hat{\sigma}_z^{(2)}
	+ J\sigma_z^{(1)}\sigma_
z^{(1)}
\end{equation}
with $J\ll \omega_1,\omega_2$,
and thus
\begin{equation}
\rho_\epsilon\approx\frac{\one}{4}+\frac{\epsilon}{4}(\hat{\sigma}_z^{(1)}+\mu\hat{\sigma}_z
^{(
2)})
\label{ps}
\end{equation}
where $\mu$ is the ratio of the Larmor frequencies of spin~$2$ and~$1$.
By a carefully tailored sequence of RF pulses, any two spin unitary
transformation of this state can be achieved.
Furthermore, using a spatially non-uniform magnetic field pulse (a gradient
pulse)  and averaging the varying phases over the
sample we can effect particular non-unitary
transformations\cite{cory:qc2000a}. With
these two techniques it is possible to prepare the
system in a so called {\it pseudo-pure state} of the form
\begin{equation}
\rho_\epsilon\approx\frac{(1-\epsilon)}{2^N}\one
	+\epsilon\vert\Psi\rangle\langle|\Psi\vert
\end{equation}
for~$|\Psi\rangle\langle\Psi|$ a pure, and possibly entangled, state
for $N$~spins. It is possible to place
bounds on the value of $\epsilon$ for which the total state in Eq.~(\ref{ps})
is entangled\cite{braunstein:qc1999a}, that is a state
which cannot be written as a convex combination of factorisable  density
operators. In typical experiments
$\epsilon\approx 10^{-5}$, a value which is too small for these states
to be entangled.  The  pseudo-pure states produced in two qubit NMR
quantum information
processing experiments are not entangled and thus the spin--spin
correlations at a fixed
time have a purely classical interpretation.

Even though entangled states have not yet been produced in NMR quantum
information processing experiments, this does not mean that the system
is not quantum mechanical. The important question is whether there is a
classical description of the dynamics of these experiments.  This is a
question that can and must be answered in a way that does not depend on
the initial and final states of the system. It is a question concerning
the propagator, or Greens function, for the dynamics, not the initial
and final  states. It is possible that the initial and final states may
exhibit no quantum correlations and have a perfectly valid classical
description, yet not be connected by a classical dynamical model, be it
deterministic or stochastic. In the case of stochastic dynamics the
answer to this question will involve a specification of transition
probabilities. As we show below, via a particular but well motivated classical model, 
it is possible that the initial and
final states are described by separable states (and thus could be
interpreted as a perfectly valid classical probability distribution)
yet no positive transition probability exist to connect them either
globally or infinitesimally.

The quantum dynamics of a single spin--half system, described by the
Hamiltonian Eq.~(\ref{single_ham}), is
equivalent to the classical dynamics of the corresponding magnetic dipole
in an applied field\cite{ernst:qc1994a,schack:qc1999a}. In both cases we find
a linear precession of the average magnetic dipole about the direction of
the applied field. Of course the states involved in the
classical and quantum case are quite different. In the quantum case we can
express the time--dependent state of the spin as
\begin{equation}
\rho(t)=\frac{1}{2}(1+\vec{n}(t)\cdot\vec{\sigma})
\end{equation}
with $\vec{n}(t)$ a unit vector and
$\vec{\sigma}=\vec{e}_x\hat{\sigma}_x+\vec{e}_y\hat{\sigma}_y+
\vec{e}_z\hat{\sigma}_z$. If the quantum Hamiltonian is
$\hat{H}=\vec{B}\cdot\vec{\sigma}$ the quantum dynamics is given by the
solution to the Bloch equations:
\begin{equation}
\frac{d\vec{n}}{dt} = -\vec{B}\times\vec{n}
\label{precess}
\end{equation}
This is the same equation of motion as a classical (unit)  magnetic dipole,
$\vec{n}$, in a magnetic field $\vec{B}$. The equation describes the linear
precession
of a point, $\vec{n}$, on the  unit sphere  around the direction of
$\vec{B}$ at a rate $|B|$.

Instead of a single classical dipole  suppose we had a distribution  of
dipoles described by some initial probability distribution function on the
sphere,
$Q_0(\vec{n})$. As the precession on the sphere is linear, each vector will
precess at the same rate $|B|$ around the direction of $\vec{B}$. The
distribution at
time is then simply $Q_0(\vec{n}(t))$. In other words the solution of
Eq.~(\ref{precess}) are the characteristic equations for the equation of
motion of the
distribution function. The distribution simply rotates without distortion
at a constant rate around $\vec{B}$.

However, the classical and quantum dynamics that result for two magnetic
dipoles interacting via the spin--spin
interaction Eq.~(\ref{two_ham}), an entangling interaction, are very different as we show below. Thus we
conclude that, while at present liquid--state NMR may not
have access to entangled quantum  states, it does allow us to
realize a quantum dynamics for those states that
is not be realized classically. It is the dynamics that is quantum in
liquid--state NMR not the states. Liquid state NMR allows
us to experimentally study the quantum dynamics of many coupled qubits and
at present
probably the most interesting element is to understand the amount of
control we have on this dynamics. The corresponding
classical system, with the same Hamiltonian, could never achieve this.

To explain this we first note an equivalence between the spin--spin
interaction of Eq.~(\ref{two_ham}) and a nonlinear
top model. Consider the collective angular momentum operator $\hat{S}_z$
defined by
\begin{equation}
\hat{S}_z=\frac{1}{2}(\hat{\sigma}_z^{(1)}+\hat{\sigma}_z^{(2)}).
\end{equation}
It is then easy to see that
\begin{equation}
\hat{S}_z^2=\frac{1}{2}\left(1+\hat{\sigma}_z^{(1)}\hat{\sigma}_z^{(2)}\right)~.
\end{equation}
Thus up to an irrelevant additive constant we obtain
\begin{equation}
H_2\equiv\frac{J}{2}\hat{S}_z^2.
\end{equation}
The question of the dynamics is reduced to studying the quantum and
classical dynamics of this
nonlinear top. Note that this Hamiltonian commutes with $\hat{S}^2$ the
total angular momentum operator. Thus the system cannot
evolve out of the subspaces corresponding to the irreducible
representations of a two spin system. There are two such
subspaces, the triplet with $s=1$ and the singlet with $s=0$. If we begin
with the state in which both spins are down we
cannot leave the triplet subspace with this Hamiltonian. Of course
combinations of the two--spin and single-spin unitary
operators will mix the two irreducible subspaces. However, the quantum and
classical dynamics that results from the nonlinear
top interaction (with the same hamiltonian form),  equivalently the two--spin interaction, are different
regardless of the initial states as we now show. To be
specific we will consider the dynamics restricted to the triplet  ($s=1$)
subspace.

We will follow closely the presentation of Sanders\cite{Sanders89} concerning
the classical and quantum dynamics of
nonlinear tops. We will assume that the physical interaction between two
spins is fixed as the scalar coupling of two
magnetic dipoles. The corresponding Hamiltonian is then fixed and we can
compare the dynamics of observable quantities that
results when the interaction is treated either quantum mechanically or
classically.  It is conceivable that the exact quantum
dynamics could be simulated exactly by  a {\it different}  classical
Hamiltonian. After all we could always
simulate the quantum system on a classical computer which is indeed  a
classical system with a very complex time--dependent
Hamiltonian.  However, we believe it is unlikely that any classical
Hamiltonian no matter how complex could simulate the quantum
dynamics over a {\it fixed} time interval. We will return to this point in
the discussion section below.

The classical dynamics of a
nonlinear top is defined by the Hamiltonian
\begin{equation}
H=\omega S_z+\frac{J}{2s} S_z^2~,
\end{equation}
where we have included a linear precession term with~$\omega$ the linear
precession frequency.
In this case the quantity $S_z$ is the z-component of the
classical angular
momentum of the top. The first term describes the
linear precession of the angular momentum vector about the $z$ axis at the
constant rate $\omega$. The second term
describes a nonlinear precession about the $z$ axis at a frequency that
depends on the $z$ component of angular momentum.
The classical mechanics is
described by the motion of a point in a spherical  phase-space embedded in
the three--dimensional Euclidean space with
coordinates $S_x,S_y,S_z$ with $S_x^2+S_y^2+S_z^2=s^2\one$~\cite{Lassig93}. The
classical states
are probability distributions which describes an ensemble of tops with a
distribution of  angular momentum directions
(every top in the ensemble has the same magnitude of total angular
momentum). The  points on the sphere of
radius $s$ are conveniently parameterized in polar coordinates as
\begin{equation}
{\bf S}/s=(\sin\theta\cos\phi, \sin\theta\sin\phi, \cos\theta)
\end{equation}
However, we will use the stereo-graphic projection of the sphere onto the
complex plane  defined by the map
\begin{equation}
z=e^{i\phi}\tan\theta/2
\end{equation}
The north pole ($S_z=s$) is mapped to the origin ($z=0$) and the south pole
is mapped to infinity. The equator is mapped
to the unit circle.  In this conformal mapping, distributions with circular
contours are mapped to distributions with circular
contours in the complex plane. The dynamics of a distribution of points is
easily described. A linear precession about the $z$
axis simply causes a distribution of point to rotate about the origin in
the complex plane without changing it shape. However
the nonlinear precession causes the distribution to shear as different
parts of the distribution with different values for $S_z$
may have different precession rates. In the long time limit the
distribution will tend to become smeared around the origin
in the complex plane (see Sanders\cite{Sanders89} for a pictorial
representation). As we will see this is very different from
what happens in the quantum case where the shearing ceases after some time
and revivals and fractional revivals of the
initial state occur.

In order to make a comparison with the quantum dynamics we need to consider
the dynamics of a distribution of points on the
sphere. This is because a quantum state cannot be perfectly localized at a
point on the sphere.  The classical state of the
system is described by a probability distribution
$Q(z)$ of the vectors
$\bf S$ corresponding to $z$. The expectation values for the components of
angular momentum are given by
\begin{eqnarray}
{\cal E}(S_z)& = & \int d \mu(z)Q(z)\frac{s(1-|z|^2)}{1+|z|^2}\\
{\cal E}(S_x-iS_y)  & = & \int d \mu(z) Q(z)\frac{2sz^*}{1+|z|^2}
\end{eqnarray}
where the integration measure in the stereo-graphic plane is
\begin{equation}
d \mu(z)=\frac{2s+1}{\pi}(1+|z|^2)^{-2}~.
\end{equation}
We have chosen the  prefactor $2s+1$  as a scaling of the classical
probability distribution, which makes the
comparison with the quantum case more convenient.

The classical dynamics is described by a Liouville
equation
\begin{equation}
\frac{\partial Q}{\partial t}=\{H,Q\}
\end{equation}
where the Poisson bracket $\{\ ,\ \}$ may be determined using
$\{S_i,S_j\}=\sum_k\epsilon_{ijk}S_k$. The Liouville equation
is a first--order partial differential equation of the form
\begin{equation}
\frac{\partial Q(z,t)}{\partial t}=-\dot{z}\ \frac{\partial \ }{\partial
z}Q(z,t)+\mbox{c.c}
\label{c_evol}
\end{equation}
where the equations of motion are
\begin{equation}
\dot{z}=i\left (\omega+J\frac{1-|z|^2}{1+|z|^2}\right ) z\ .
\end{equation}
The solution to this is easily found after noting that $|z|^2$ is a
constant of motion. Thus
\begin{equation}
z(t)=\exp\left [-it\left (\omega+J\frac{1-|z|^2}{1+|z|^2}\right )\right
]z(0)\ ,
\end{equation}
which in polar coordinates becomes,
\begin{eqnarray}
\theta(t) & = & \theta_0\\
\phi(t)  & = & \phi_0-\omega t-Jt\cos\theta_0\ .
\end{eqnarray}
where $\phi_0,\ \theta_0$ are the initial values.  In this form it is
particularly easy to see that the dynamics is a rotational shear of the
sphere around the~$z$ axis.

The solution for the probability density is
\begin{equation}
Q(z,t)=\int d^2z^\prime{\cal T}(z,z^\prime;t)Q(z^\prime,0)
\label{class_prop}
\end{equation}
where the propagator is defined by
\begin{equation}
{\cal T}(z,z^\prime;t)=\delta^2(z^\prime(t)-z)
\end{equation}
where $z^\prime(t)$ is the solution to the equation of motion for time $t$
starting with the initial point $z^\prime$. When $J=0$ we recover the
previous result for a classical magnetic dipole: the distribution simply
rotates, without distortion, at a constant rate $\omega$ about the $z$
axis. The effect of
the nonlinear term  proportional to $J$ causes a rotational shearing of the
distribution around the $z$ axis.

We may include additional stochastic dynamics on top of the Hamiltonian
dynamics. However, it is important to note that if
$Q(z,t)$ is a probability distribution then the propagators ${\cal T}$ must
be positive and may be interpreted as transition
probabilities. The Hamiltonian Liouville evolution is a special case. As
the propagator is simply the Greens function for the
evolution equation the positivity requirement for the propagator restricts
the allowed form of evolution equations. It is well
known that the allowed forms correspond to Fokker-Planck equations and can
contain at most second--order derivatives with
positive definite diffusion matrices\cite{Gardiner83}.  In other words, if
the propagators are to be positive the evolution
equation is necessarily restricted regardless of the initial or final
conditions. We may thus define allowed classical
dynamics either in terms of positive transition probabilities or in terms
of the differential operator for the dynamics.

To compare the quantum and classical dynamics we now need to define a
relevant quantum distribution. It is argued in
references \cite{Sanders89,Milburn86} that the appropriate object is the
matrix elements of the quantum density operator in a coherent state
basis. In the case of the harmonic oscillator, these are the coherent
states of the Heisenberg-Weyl group, and the resulting distribution is a
true (ie positive) probability distribution for simultaneous measurement of position
and momentum\cite{Braunstein91}. In the case of angular momentum we can
use the SU(2) coherent states\cite{Perelomov} defined by
\begin{eqnarray}
|z\rangle & = & R(z)|s,s\rangle_z\\
&  = & (1+|z|^2)^{-s}\sum_{m=0}^{2s} {2s \choose m}^{1/2}z^m|s,s-m\rangle_z
\end{eqnarray}
where $|s,m\rangle$ are the $2s+1$ eigenstates of $\hat{S}_z$, and the
rotation operator is
\begin{eqnarray}
\hat{R}(z) & = & \exp\left [-i\theta{\bf n}.\hat{\bf S}\right ]
\end{eqnarray}
with the unit vector ${\bf n}=(\sin\phi,\cos\phi,0)$.  The states
$|z\rangle$ are
product states in terms of the qubits which are rotated from the state
$|0\rangle_1|0\rangle_2$ by the angle $\phi$ and $\theta$  on the Block sphere:
$$
|z\rangle={1\over (1+|z|^2)} (|0\rangle_1 +z|1\rangle_1)\otimes(|0\rangle_2
+z|1\rangle_2)
$$
The function
\begin{equation}
Q(z,t)=\mbox{tr}(\rho(t)|z\rangle\langle z|)
\end{equation}
is a true (that is positive) probability distribution for measurements
defined by the projection operator valued measure
(POVM)
$|z\rangle\langle z|d^2z/2\pi$.  Note that all allowed distributions are
necessarily positive (and bounded) from the
construction of
$Q(z,t)$ as a trace of the product of a positive operator and a projection
operator. For example, the Q function for a particular atomic coherent
state
$|z_0\rangle$ is
\begin{equation}
Q(z) = \left [\frac{(1+z_0^*z)(1+z_0z^*)}{(1+z_0z_0^*)(1+zz^*)}\right ]^{2s}~.
\label{initQ}
\end{equation}

The first moments are given by integrals over the Q-function as
\begin{eqnarray}
\langle\hat{S}_x-i\hat{S}_y\rangle & = & \int
d\mu(z)Q(z)\frac{2(s+1)z^*}{1+|z|^2}\\
\langle\hat{S}_z\rangle & = & \int d\mu(z)Q(z)\left
(\frac{(s+1)(1-|z|^2)}{1+|z|^2}\right )
\end{eqnarray}
The important point to notice is that the averages of $\hat{S}_\pm$
are given by the same functional form as the
classical case, apart from an additional term that becomes negligible in
the semi-classical limit as
$s\rightarrow\infty$.  The second--order moments
are given by
\begin{eqnarray}
\langle \hat{S}_-^2\rangle & = &\int
d\mu(z)Q(z)\left
(\frac{2(2s+3)(s+1){z^*}^2}{(1+|z|^2)^2}\right) \\
\langle\hat{S}_z^2\rangle & = &\int
d\mu(z)Q(z)	\nonumber	\\	&&	\times
\left(\frac{(s+1)^2 -2s(s+2)|z|^2+(s^2+4j+5)|z|^4}{(1+|z|^2)^2}\right)
\end{eqnarray}

Thus, even though the Q function is a true probability distribution, its
marginals do not give the
quantum expectation values: an additional rule is needed to connect
averages over the Q function to the quantum averages.
This is analogous to the case for the harmonic oscillator coherent
states\cite{Milburn86}. In the case
of a spherical phase space, however, the difference appears already at the
level of the first--order moments.

Taking matrix elements of the quantum Liouville equation
\begin{equation}
\frac{d\rho}{dt}=-i[H,\rho]~,
\end{equation}
we obtain the evolution equation
\begin{eqnarray}
\frac{\partial\ }{\partial t}Q(z,t)
	&=& -i\left (\omega
+J\frac{1-|z|^2}{1+|z|^2}-\frac{J}{2s}z\frac{\partial \
}{\partial z}\right)	\nonumber	\\
	&& \times z\frac{\partial \ }{\partial z}Q(z,t)+\mbox{c.c}
\label{q_evol}
\end{eqnarray}
This equation is linear in $Q$; thus, $\rho_1$ in Eq.~(\ref{thermal}) will obey
exactly the same equation.
In the limit of $s\rightarrow\infty$, with $\omega,\lambda$ held constant,
the equation reduces to the  first--order differential
equation of classical dynamics. The difference between quantum and
classical dynamics is due to the second
order differential operators. Note that  while these terms are second
order, they are certainly not of the kind expected for a diffusion
equation, as the
corresponding diffusion matrix would not be positive definite. This is a
familiar feature of the difference between classical and quantum dynamics
as reflected in
the  dynamics of a quasi probability distribution and was first noted in
the context of quantum optics\cite{WallsMilbun}

In reference \cite{Sanders89} it is was shown how extreme this difference
could be. For
example, at times $t=\pi s/J$ an initial coherent state (or coherent
pseudo-pure state)
$|z_0\rangle$ would evolve, in the rotating frame ($\omega=0$) into the
pure state
\begin{equation}
2^{-1/2}\left (e^{-i\pi/4}|z_0\rangle+(-1)^se^{i\pi/4}|-z_0\rangle\right )
\end{equation}
for which the resulting Q-function is double peaked.
The state is entangled in terms of the pure state
representation.  No state such as this could ever be obtained from the
classical dynamics given in Eq.~(\ref{class_prop}) starting from the
initial state Eq.~(\ref{initQ}).


\section{Discussion}
We noted above that the difference between the quantum and classical
dynamics  appears through the
second--order derivatives in Eq.~(\ref{q_evol}). How does this difference
become manifest in the propagator for the equation?
It is far from clear that we can find a positive  propagator corresponding
to the differential operator on the right
hand side of Eq.~(\ref{q_evol}), especially as this equation is not  of
Fokker-Planck form (the diffusion matrix is not positive
definite). Nonetheless all initial and final Q-functions must be positive
by construction.  We now show that the quantum
evolution of the Q-function cannot be written in terms of transition
probability propagators.
Despite this it can be obtained uniquely from the initial condition, $Q(z,t=0)$

To begin we define the
Q-function {\em amplitude} for a pure state
$|\psi(t)\rangle$
\begin{equation}
{\cal Q}(z,t)=\langle z|\psi(t)\rangle
\end{equation}
The Q-function is given by the modulus squared of the Q-function amplitude,
$Q(z,t)=|{\cal Q}(z,t)|^2$. The linearity of the
Schr\"{o}dinger equation now requires that the dynamics of the Q-amplitude
can be written in terms of a linear propagator
\begin{equation}
{\cal Q}(z,t)=\int d^2z^\prime{\cal L}(z,z^\prime;t){\cal Q}(z^\prime,0)
\end{equation}
where ${\cal Q}(z,0)$ is the Q-function amplitude for the initial state and
the propagator (for $s=1$) is constructed from the unitary
time evolution operator
$U(t)$ as
\begin{eqnarray}
{\cal L}(z,z_1;t) &=&\langle z|U(t)|z_1\rangle\\
 &=&\frac{1}{(1+|z|^2)(1+|z_1|^2)}
\Bigg( e^{-i(\omega +\frac{J}{2s})t} + 2z^*z_1	\nonumber	\\
	&& +{z^*}^2 {z_1}^2
e^{-i(-\omega +\frac{J}{2s})t}  \Bigg)
\end{eqnarray}
Thus the Q-function dynamics can be written in integral form as
\begin{eqnarray}
Q(z,t)&=&\int d^2z_1 d^2z_2{\cal L}(z,z_1;t){\cal L}^*(z,z_2;t)
		\nonumber	\\	&&\times
	{\cal Q}(z_1,0)^*{\cal Q}(z_2,0)
\end{eqnarray}
In the case of a general initial state $\rho(0)$ the equation becomes
\begin{equation}
Q(z,t)=\int d^2z_1 d^2z_2{\cal L}(z,z_1;t){\cal L}^*(z,z_2;t)\langle
z_1|\rho(0)|z_2\rangle
\label{Q_prop}
\end{equation}

This last expression seems to suggest we need to know more than just the
initial condition $Q(z,0)$, but this is not the case.
The matrix elements of $\rho$ in the coherent state basis suffice to
uniquely determine the state and thus uniquely determine the off--diagonal
matrix
elements\cite{Perelomov}, by analytic continuation.  A similar statement
may be made about the propagator in Eq.~(\ref{Q_prop}). We only need to know
the diagonal matrix element of $U^\dagger|z\rangle\langle z|U$ in order to
determine the total propagator in Eq.~(\ref{Q_prop}).
Thus knowledge of the initial Q-function $Q(z,0)$ and a positive linear
propagator
${\cal K}(z,z_1;t)=\langle z_1|U(t)^\dagger|z\rangle\langle
z|U(t)|z_1\rangle$ uniquely determine the solution to the quantum
evolution equation for the Q-function. This is of course also true for the
classical evolution equation, through a simple convolution of the
initial state and the propagator. However, the propagation integral in the
quantum case, Eq.~(\ref{Q_prop}), is not a positive (or even real) function
and can
have no interpretation as a transition probability.

It is generally accepted that uncontrolled interactions with an environment
enable the quantum
and classical dynamics to be reconciled when states of the environment are
averaged over\cite{zurek}.
In the case of two coupled spins, there are a variety of possible
environmental interactions. In order to
illustrate the principle of decoherence, we take the simplest possible case
in which both spins are coupled equally to the
same environment with a Hamiltonian that conserves the total z--component of
angular momentum, $\hat{S}_z$.
While this collective dephasing model is not very realistic for NMR
experiments it will illustrate how decoherence can cause the
quantum propagator in Eq.~(\ref{Q_prop}) to become diagonal.

The collective dephasing master equation is given by\cite{Sanders89}
\begin{equation}
\dot{\rho}={\cal
D}\rho=-i[H,\rho]-\frac{\gamma}{2s}[\hat{S}_z,[\hat{S}_z,\rho]]
\end{equation}
The general solution for the Q-function may then be written as
\begin{equation}
Q(z,t)=\int d^2z_1\int d^2z_2{\cal P}(z;z_1,z_2,t)\langle
z_1|\rho(0)|z_2\rangle
\end{equation}
where the propagator is given in terms of the coherent state matrix
elements of the dynamically propagated off--diagonal projector,
\begin{equation}
{\cal P}(z;z_1,z_2,t)=\langle z|e^{{\cal D}t}\left (|z_1\rangle\langle
z_2|\right )|z\rangle
\end{equation}
For short times we can expand this to linear order in $t$. The dominant
non--Hamiltonian terms in total spin $s$ are given by
\begin{eqnarray}
{\cal P}(z;z_1,z_2,t) &=& \left [1-\frac{\gamma  s t}{2}\left
(\frac{|z_1|^2-|z_2|^2}{(1+|z_1|^2)(1+|z_2|^2)}\right )^2\right ]
	\nonumber	\\
	&&\times {\cal P}(z;z_1,z_2,0)+\cdots
\end{eqnarray}
The cofactor of $\gamma s t/2$ is less than unity.
For two spin half systems with $s=1$ the coherence decay rate is then set entirely by
$\gamma$ which could be small
compared to the time scale set by the cohernet interaction, $J^{-1}$.
When the initial state is concentrated on $|z_1|=|z_2|$ 
the decoherence is also small. 
The coherence decay rate for large semiclassical systems for which $s>>1$
can be large.  Of course the resulting
propagator is not the same as the classical Hamiltonian result
(Eq.~(\ref{class_prop})) in the absence of  dephasing
as the coupling to the environment would add some level of
phase diffusion to the classical dynamics as well, broadening the delta
function in Eq.~(\ref{class_prop}).
In the case of $s=1$ considered here one would need to consider all the
terms in the short time expansion for ${\cal P}(z;z_1,z_2,t)$,
but it remains the case that terms with large values of $|z_1|-|z_2|$ are
rapidly suppressed.

Other two--qubit interactions might be considered, for example the exchange
interaction\cite{divincenzo:qc2000a},
\begin{equation}
H_{\rm ex} = \frac{J}{4}\vec{\sigma}^{(1)}\cdot\vec{\sigma}^{(2)}
   = \frac{J}{2}(\vec{S}\cdot\vec{S}-\frac{3}{2})
\end{equation}
where the total spin operator is
$\vec{S}=(\vec{\sigma}^{(1)}+\vec{\sigma}^{(2)})/2$.
This is the Hamiltonian for a rigid body wiht an isotropic moment of intertia 
with no external torques and with moment of inertia $J^{-1}$. The
classical phase space is a cylinder with canonical coordinates $S$ on the axis
of the cylinder and $\theta$
around the circumference. The classical equations
are the same as those of a free mass with periodic boundary conditions in
position at $\pm\pi$ : $\dot{\theta}=Js$ and
$\dot{S}=0$. The first of these again indicates a rotational shearing of a
localized distribution on the phase space. When the shearing
causes different parts of the state to overlap on the classical phase
space, we expect the corresponding quantum system to exhibit
interference fringes. Thus the quantum and classical dynamics of two
interacting magnetic dipoles with such an interaction must differ. (Clearly
such an
interaction would need a different mechanism in classical systems as
exchange interactions are quantum mechanical.)

As we have seen the nonlinear top can generate the superposition state
$(e^{-i\pi/4}|s,s\rangle_z+e^{-i\pi/4}|s,-s\rangle_z)$. A similar state can
be generated
by a sequence of CNOT gates on $N=\log_2(2j+1)$ qubits. A product state of
$N$ qubits can be written in terms of a binary string
$|X\rangle=\Pi_i^\otimes|x_i\rangle$, where $x_i$ is the i'th term of the
string $X$. Alternatively $X$ could encode an integer $k$ in binary form.
The maximally
entangled state of $N$ qubits, $|000\ldots 0\rangle+|111\ldots 1\rangle$
would then take the form $|0\rangle+|M\rangle$ where $M=2^N$. Such as state
is easily
generated in a quantum computer by a single Hadamard gate on the first
qubit followed by a cascaded sequence of controlled NOT gates.
If we change the notation for angular momentum
states so that $|s,m-s\rangle_z=|m\rangle$ where
$m=0,1,\ldots 2s$, then the angular momentum superposition state generated
by the nonlinear top is equivalent to the maximally entangled state. This
equivalence
suggests that a nonlinear top may itself be made to act as a quantum
computer in the $2s+1$ dimensional Hilbert space. Indeed numerical evidence
exists\cite{Xu99}
that a time--dependent Hamiltonian of the form
$H(t)=\vec{B}(t)\cdot\vec{S}+J(t)S_z^2$ can generate any state in the
Hilbert space by a suitable choice of the time
dependent coefficients. We conjecture that a sequence of pulses
$U(\theta_n,\phi_n,\chi_n)=\exp(-i\theta_n\vec{n}_n\cdot\vec{S})\exp(-i\chi_nS_z
^2)$ where
$\vec{n}=(\cos\phi,\sin\phi,0)$ will suffice to perform quantum algorithms
in the Hilbert space of the nonlinear top.
Further work remains to be done
to determine whether there are interesting efficient algorithms
in terms of number of pulses and (say) $\log(s)$.

\section{Conclusion}

We have investigated classical and quantum Hamiltonians of the non linear
top and show that they produce different observables even in the presence
of highly mixed states.  This shows that even if a state is separable its
classical and quantum evolution can differ.  Thus this fortifies the claim that
the evolution is one of the quantum elements of liquid--state NMR QIP. 
However it does not prove that bulk ensmeble NMR quantum computation is uniquely quantum. 
We have used a particular classical model, that is one 
motivated by the physics of two interacting magnetic dipoles, to compare the quantum and classical dynamics. 
It may well be the  case that
for states that are close to the maximally mixed state there is another 
classical model (in effect a hidden variable model)
that correctly describes the dynamics as far as is required to model the observed 
results in NMR experiments\cite{schack:qc1999a}. 

Even if the transformation of the state is not classical our present
work does not show that using these mixed state qubits is as powerful
as a pure state quantum computer. If we use pseudo-pure state as has
been done in present experiments, the signal to noise decreases
exponentially, thus rendering the algorithms inefficient. In this case
liquid--state NMR does not provide any advantage (with respect to
speed) over classical computers. In the absence of noise, Schulman and
Vazirani\cite{schulman:qc1998a} have shown how to efficiently
transform the initial mixed state into a pure state.  Some simulations
of quantum systems evolving under unitary transform and the algorithm
suggested in \cite{knill:qc1998a} could also be implemented
efficiently as long as noise is negligible. Thus liquid--state NMR (in
the absence of noise, i.e. on time scales smaller than $T_2$) offers a
testing ground for QIP.  The dynamics is the correct one.  Moreover it
has the right error model to provide a test bed for small quantum
computations (up to roughly 10 qubits). The difference between liquid
state NMR and other proposals for quantum computation is that, in the
former case we do not have, even in principle, a method to make the
quantum evolution robust.  In practice this does not distinguish NMR
from other present devices as none have approached the accuracy needed
for scalability.

In the present work we have neglected the possibility of hidden
variables. The reason is that there is a trivial hidden variable model
which would explain all the present experiments in QIP (including not
only NMR but also other technologies) as long as observations are not
made on space-like surfaces.  The model can be thought of as a
classical computer which simulates the quantum
evolution and tells the bits of the physical system how to behave so
they mimic quantum mechanics.  Although this model can describe all
the experiments in QIP today, the amount of resources it uses compared
to its quantum counterpart always seem to be exponential in the size
of the problem (the model of Schack and Caves is such a model).  It is
important that the resources account for not only the signal to noise
but all resources as it is usually easy to trade one resource for
another.  We would like to conjecture that what distinguishes
classical and quantum computation is the amount of algorithmic
information required to produce the answer of some problem starting
from a fiducial state.  Classical devices require an
exponential amount of algorithmic information compared to a quantum
computer to answer certain problems.  Simulating liquid--state NMR
experiments is one such problem compared to known classical algorithms.

So where did the power of quantum computation come from? In this paper
we have argued that the power comes from the dynamics of the
system. A similar argument has been made by others\cite{braunstein:qc1999a,schack:qc1999a}.
We have criticized the view that entanglement is {\it the}
source of power of quantum computation by giving algorithms and
dynamical evolution which do not depend on entanglement.  There may be
a variety of elements which make these devices more powerful, and for
another unusual method to quantum compute see \cite{knill:qc2000a}.
However, as long as we lack a proof that we cannot simulate quantum
systems efficiently, it is hard to attribute the source of the power
of quantumness, and we await more powerful arguments.

\section{Acknowledgment}

GJM gratefuly acknowledges discussions with Carl Caves. EK and RL acknowledge support from the US DOE under contract W-7405-ENG-36.
Much of this work was completed while GJM, RL and EK were at the  Aspen Center for Physics.
\end{multicols}


\begin{references}
\bibitem{mixed-state} We define a completely mixed state as one for which the 
von Neumann entropy is a maximum. 
A highly mixed state is one for which the entropy is close to the maximum.
\end{references}

\newpage

\begin{figure}
\includegraphics[width=12cm,height=9cm]{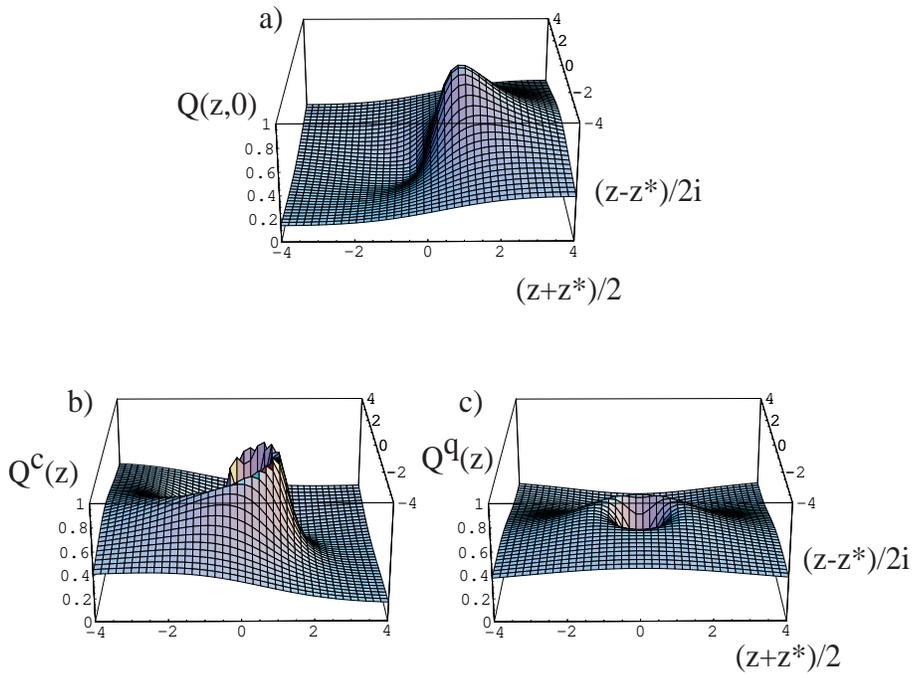}
\caption{Comparison between the classical and quantum evolution of the Q(z)
function.  The initial state is given by a).  It corresponds to the
distribution for the quantum state $|z=1\rangle$.  b) and c) depict the
classical and quantum distribution at time $t=2\pi/J$ (for $\omega=0$).
The classical evolution follows the equation of motion Eq.~(\ref{c_evol})
and the quantum one Eq.~(\ref{q_evol}).
The discrepancy between the classical evolution and the
quantum one is evident.}
\label{figure:fig1}
\end{figure}

\end{document}